\documentclass[10pt,letterpaper]{article}
\usepackage{color}
\usepackage[colorlinks=true,urlcolor=blue,linkcolor=black,citecolor=black,breaklinks=true,bookmarks=false]{hyperref}
\usepackage{graphicx}
\usepackage{dcolumn}
\usepackage{bm}

\begin{document}

\title{Resonance interaction of two dipoles in optically active surroundings}
\author{Qi-Zhang Yuan$^1$, Chun-Hua Yuan$^{1,*}$ and Weiping Zhang$^1$}
\maketitle

{$^1$State Key Laboratory of Precision Spectroscopy, Department of
Physics, East China Normal University, 500 Dongchuan Road Shanghai 200241,
People's Republic of China.}

{$^*$chyuan@phy.ecnu.edu.com}

\begin{abstract}
We study the resonance interaction between two quantum electric dipoles
immersed in optically active surroundings. Quantum electrodynamics is
employed to deal with dipole-vacuum interaction. Our results show that the
optical activity of surroundings will not change the single atom behaviors
while it can change the collective behaviors of the two dipoles, as well as
greatly affect the dipole-dipole resonance interaction. Especially, if the
orientations of two dipoles are orthogonal and respectively perpendicular to
the interdipole axis, the interdipole resonance interaction can be
established with the help of optically active surroundings while there is no
resonance interaction in vacuum.
\end{abstract}



\section{Introduction}

Resonance interaction (RI)~\cite{ri2,ri3} is a mechanism describing two
quantum emitters (one is excited and the other one is in ground state) with
same transition frequencies interact with each other by exchanging virtual
photons. RI represents the nature of the interaction between vacuum
electromagnetic field and atomic system, and also plays important roles in
broad application fields. For instance, RI can contribute to operation of
solid state lasers and fluorescent tubes~\cite{ncomms4610}. In quantum
physics, RI is very useful in creating entanglement~\cite%
{entanmole,entanquad}, and also plays important roles in laser cooling~\cite%
{lasercool}, creating cold molecules~\cite{moleform1,moleform2} and quantum
storage~\cite{qs1,qs2,qs3}. The study of RI can contribute in developing
precise measurement methods in chemistry and biology for metering distance
between molecules ~\cite{moledis} or between domains in one protein which
can provide information about protein conformation ~\cite{tufe}, so it
attracts the interests of chemists and biologists as well as physicists.
Further more, RI between quantum emitters is the essence of F\"{o}rster
resonance energy transfer (FRET)~\cite{dd,forster}, while the latter one
lies at the core of biophysics cause it represents the nature of
intermolecule interaction, provides the foundation of several detection
methods~\cite{ugfeb,fretmi} and plays important role in biological phenomena
such as photosynthesis. The investigation into photosynthesis can help in
increasing the efficiencies of solar cells (specially the dye-sensitized
solar cells~\cite{dssc}), which is in favor of the usage of solar energy.

In biological phenomena, RI generally takes place in living cells filled
with solution of organic compounds, which are optically active in most
cases. Similarly, the electrolyte solution used in dye-sensitized solar
cells contains volatile organic compounds~\cite{ysm} and leads to an
optically active surroundings. When the dipoles (which can be atoms,
molecules, or just chromophores and fluorophores) are immersed in optically
active solutions, the interaction between them must be affected by the
optically active surroundings. The study of this effect can help in
understanding the processes in natural photosynthesis, while the simulation
of which can contribute in improving the efficiency of solar cells. We think
this effect is important as well as interesting, but so far as we all know,
there is an absence of detailed theoretical investigation of this effect.

In this paper, we study the RI between two quantum dipoles which are
immersed in optically active surroundings. Quantum electrodynamics is
employed to deal with dipole-vacuum interaction. We give out the analytical
expressions of interdipole RI in optically active surroundings as well as
several numerical examples. We find that the collective behaviors of the
dipole pair (containing both the collective radiation rates and the
interaction induced level shifts) are affected by optically active
surroundings while the single atom properties are not. When the
orientations of two dipoles are orthogonal and respectively perpendicular to
interdipole axis, the optical rotation of the surrounding medium can cause
great interaction between the dipoles while there is no interaction in
vacuum.

\section{Dynamics of two dipoles interaction}

\label{dynamic} In this section, a brief review of the dynamical analysis of
dipole-reservoir interaction will be given out.

We calculate the dipole-light interaction problem in Schr\"{o}dinger
picture. The two dipoles are marked by 1 and 2. The quantum state of the
system can be expressed by
\begin{equation}
\left\vert \psi \right\rangle =\sum_{i=1,2}C_{i}\left\vert i\right\rangle
\left\vert 0\right\rangle +\sum_{i=3,4}\sum_{\boldsymbol{k}\lambda }D_{i,%
\boldsymbol{k}\lambda }\left\vert i\right\rangle \left\vert \boldsymbol{k}%
\lambda \right\rangle ,
\end{equation}%
in which the basic states $\left\vert 1\right\rangle -\left\vert
4\right\rangle $ are
\begin{equation}
\left\vert 1\right\rangle =\left\vert e_{1}g_{2}\right\rangle,~\left\vert
2\right\rangle =\left\vert g_{1}e_{2}\right\rangle,~\left\vert
3\right\rangle =\left\vert e_{1}e_{2}\right\rangle,~\left\vert
4\right\rangle =\left\vert g_{1}g_{2}\right\rangle .
\end{equation}%
$\left\vert e\right\rangle $ and $\left\vert g\right\rangle $ are the
quantum states of the dipoles. $\left\vert g_{i}\right\rangle $ represents
that the dipole $i$ is in ground state and $\left\vert e_{j}\right\rangle $
means dipole $j$ is in excited state. For example, the ket $\left\vert
1\right\rangle =\left\vert e_{1}g_{2}\right\rangle $ represents dipole $1$
is excited while dipole $2$ is in ground state. Electromagnet field state $%
\left\vert \boldsymbol{k}\lambda \right\rangle $ means a photon with wave
vector $\boldsymbol{k}$ and polarization $\lambda $ exists, while $%
\left\vert 0\right\rangle $ represents there is no photon (the
electromagnetic vacuum state). $C_{i}$ and $D_{i,\boldsymbol{k}\lambda }$
are probability amplitudes.

The total Hamiltonian $\hat{H}$ can be departed into three parts: the
electric dipole Hamiltonian $\hat{H}_{D}$, the electromagnetic Hamiltonian $%
\hat{H}_{F}$ and the atom-light interaction Hamiltonian $\hat{H}_{INT}$. The
dipole Hamiltonian can be written as
\begin{equation}
\hat{H}_{D}=\hbar \sum_{i}\omega _{i}\hat{\sigma}_{ii},
\end{equation}%
in which $\hat{\sigma}_{ij}=\left\vert i\right\rangle \left\langle
j\right\vert $ are dipolar transition operators. The electromagnetic
Hamiltonian is
\begin{equation}
\hat{H}_{F}\left( \boldsymbol{r}\right) =\sum_{\boldsymbol{k}\lambda }\hbar
\omega _{k}\left( \hat{a}_{\boldsymbol{k}\lambda }^{\dag }\hat{a}_{%
\boldsymbol{k}\lambda }+\frac{1}{2}\right) ,
\end{equation}%
where $\hat{a}_{\boldsymbol{k}\lambda }^{\dag }$ and $\hat{a}_{\boldsymbol{k}%
\lambda }$ are creation and annihilation operators for photon of mode $%
\boldsymbol{k}\lambda $, respectively. $\hbar $ is modified Planck constant
and $\omega _{k}$ is the angular frequency of the photon. The interaction
Hamiltonian writes
\begin{equation}
\hat{H}_{INT}=-\sum_{a=1,2}\boldsymbol{\hat{d}}_{a}\cdot \boldsymbol{\hat{E}}%
\left( \boldsymbol{r}_{a}\right) ,
\end{equation}%
where $\boldsymbol{\hat{d}_{a}}=\sum_{ij}{\boldsymbol{d}}_{ij}^{a}\hat{\sigma%
}_{ij}$ are the electric dipole operators, and ${\boldsymbol{d}}_{ij}^{a}$
are the dipole transition matrix elements. $\boldsymbol{\hat{E}}$ is the
electric field operator which is given by
\begin{equation}
\boldsymbol{\hat{E}}\left( \boldsymbol{r}\right) =\sum_{\boldsymbol{k}%
\lambda }A_{k}\hat{a}_{\boldsymbol{k}\lambda }\boldsymbol{W}\left(
\boldsymbol{k}\lambda ,\boldsymbol{r}\right) +H.c,
\end{equation}%
where $A_{k}$ is the normalization coefficients, and the electric wave mode
function $\boldsymbol{W}\left( \boldsymbol{k}\lambda ,\boldsymbol{r}\right) $
is decided by the surrounding medium and the boundary conditions. With the
initial condition that dipole 1 is excited and dipole 2 is unexcited, the
probability amplitude can be worked out:
\begin{equation}
C_{1}\left( t\right) =e^{A_{L}t}\frac{e^{A_{T}t}+e^{-A_{T}t}}{2},~C_{2}\left(
t\right) =e^{A_{L}t}\frac{e^{A_{T}t}-e^{-A_{T}t}}{2}.  \label{c1c2}
\end{equation}%
The expressions of coefficients $A_{L}$ and $A_{T}$ as well as calculation
process are given in appendix A. Then we further define dipole-exchanging
symmetric (DES) state $\left\vert +\right\rangle $ and dipole-exchanging
anti-symmetric (DEAS) state $\left\vert -\right\rangle $ which is given by
\begin{equation}
\left\vert \pm \right\rangle =\frac{1}{\sqrt{2}}\left( \left\vert
e_{1}g_{2}\right\rangle \pm \left\vert g_{1}e_{2}\right\rangle \right) ,\label{symstate}
\end{equation}%
and the corresponding probability amplitudes are
\begin{equation}
C_{\pm }=\frac{1}{\sqrt{2}}\left( C_{1}\pm C_{2}\right) =\frac{1}{\sqrt{2}}%
e^{\left( A_{L}\pm A_{T}\right) t}.\label{cs}
\end{equation}%
It is obviously that $\left\vert \Re \left( A_{L}\pm A_{T}\right)
\right\vert $ and $\Im \left( A_{L}\pm A_{T}\right)$ are the damping rates
and the level shifts of DES and DEAS states, respectively. $A_{L}$
refers to single dipole mechanism and $A_{T}$ represents collective
behavior. According to the solution we get above, the dipole-dipole
interaction energy can be calculated as
\begin{equation}
E_{int}=-2\hbar \Im A_{T}\left( \left\vert C_{+}\right\vert ^{2}-\left\vert
C_{-}\right\vert ^{2}\right) .  \label{Ei}
\end{equation}%
The interdipole interaction energy is proportional to the collective level
shift ($\Im A_{T}$) times the population difference between DES and DEAS
states.

\section{Electromagnetic field in optically active medium}

\label{elefield} In last section, we have given out a brief review of
dipole-vacuum interaction. The general formulas are calculated, but to
obtain the detail expressions, the mode function $\boldsymbol{W}\left(
\boldsymbol{k}\lambda ,\boldsymbol{r}\right)$ must be specified. In this
section, we will give the function $\boldsymbol{W}\left( \boldsymbol{k}%
\lambda ,\boldsymbol{r}\right)$ in optically active medium.

One beam of left (right) hand circularly polarized light with wave vector $%
\boldsymbol{k}$ can be expressed by Jones vector as
\begin{equation}
{\boldsymbol{W}}_{left/right}=\frac{1}{\sqrt{2}}\left( \boldsymbol{\hat{e}}_{%
\boldsymbol{k},1} \pm i \boldsymbol{\hat{e}}_{\boldsymbol{k},2}\right) e^{i
\boldsymbol{k\cdot r}},
\end{equation}
$\boldsymbol{\hat{e}}_{\boldsymbol{k},1}$ and $\boldsymbol{\hat{e}}_{%
\boldsymbol{k},2}$ are unit vectors which are orthogonal to each other and
perpendicular to wave vector $\boldsymbol{k}$. $i$ is the imaginary unit.
According to Fresnel's theorem, the phase velocities of left hand and right
hand polarized light is different in optically active medium, then the general
expression of light in optically active medium can be written as
\begin{equation}
\boldsymbol{W}\left( \boldsymbol{k}\lambda ,\boldsymbol{r}\right) =\frac{1}{%
\sqrt{2}} \left( \boldsymbol{\hat{e}}_{\boldsymbol{k},1}+s_{\lambda }i%
\boldsymbol{\hat{e}}_{\boldsymbol{k},2}\right) e^{in_{\lambda }\boldsymbol{%
k\cdot r}},  \label{Wge}
\end{equation}
in which $\lambda = L$, $R$ represent the left and right hand circularly
polarized components, respectively, and $s_{L}=1$, $s_{R}=-1$. $n_{\lambda}$
are the refractive indexes for the two circularly polarized components and $%
\boldsymbol{k}$ is wave vector in free space. Using Eq. (\ref{Wge}), the
expression of function $G_{aa\left(b\right)}$ in Eq. (\ref{Aij}) can be
derived
\begin{eqnarray}
G_{11}\left( \boldsymbol{k}\lambda \right) =\frac{1}{2}\boldsymbol{d}%
_{1}\cdot \mathbf{M} \left(\boldsymbol{k}\lambda\right)\cdot \boldsymbol{d}%
_{1}, ~~G_{12}\left( \boldsymbol{k}\lambda \right) =\frac{1}{2}\boldsymbol{d}%
_{1}\cdot \mathbf{M} \left(\boldsymbol{k}\lambda\right)\cdot \boldsymbol{d}%
_{2} e^{-i n_{\lambda } \boldsymbol{k} \cdot \left(\boldsymbol{r}_{1}-%
\boldsymbol{r}_{2}\right)},\\
G_{22}\left( \boldsymbol{k}\lambda \right) =\frac{1}{2}\boldsymbol{d}%
_{2}\cdot \mathbf{M} \left(\boldsymbol{k}\lambda\right)\cdot \boldsymbol{d}%
_{2}, ~~G_{21}\left( \boldsymbol{k}\lambda \right) =\frac{1}{2}\boldsymbol{d}%
_{2}\cdot \mathbf{M} \left(\boldsymbol{k}\lambda\right)\cdot \boldsymbol{d}%
_{1} e^{-i n_{\lambda } \boldsymbol{k} \cdot \left(\boldsymbol{r}_{2}-%
\boldsymbol{r}_{1}\right)},  \label{G1112}
\end{eqnarray}%
in which
\begin{equation}
\label{G1113}
\mathbf{M}\left(\boldsymbol{k}\lambda\right)= \left[
\boldsymbol{\hat{e}}_{\boldsymbol{k},1}\boldsymbol{\hat{e}}_{\boldsymbol{k}%
,1} +\boldsymbol{\hat{e}}_{\boldsymbol{k},2}\boldsymbol{\hat{e}}_{%
\boldsymbol{k},2} +s_{\lambda }i \left( \boldsymbol{\hat{e}}_{\boldsymbol{k}%
,1}\boldsymbol{\hat{e}}_{\boldsymbol{k},2} -\boldsymbol{\hat{e}}_{
\boldsymbol{k},2}\boldsymbol{\hat{e}}_{\boldsymbol{k},1} \right)%
\right].
\end{equation}

\section{Damping rates and level shifts}

\label{drls} After defining the functions $G_{ab}$, Eqs. (\ref{Aij}) can be
calculated. We shall go to the continuum limit%
\begin{equation}
\sum_{\boldsymbol{k}}\rightarrow\frac{V}{\left( 2\pi\right) ^{3}}\int d^{3}k,
\end{equation}
and make use of Eq. (\ref{G1112}) to obtain the analytical expressions of $%
A_{L}$ and $A_{T}$
\begin{eqnarray}
&&A_{L}/\Gamma _{0}=-\sum_{\lambda }\frac{n_{\lambda }}{4} +i \sum_{\lambda }%
\frac{n_{\lambda }}{2\pi } \int_{0}^{\infty }d\tilde{\xi}\frac{\tilde{\xi}%
^{4}}{\tilde{\xi}^{2}-1},  \label{al} \\
&&A_{T}/\Gamma _{0}=-F_{1}\left( R\right) +i F_{2}\left( R \right),
\label{at}
\end{eqnarray}%
in which
\begin{eqnarray}
F_{1}\left( R\right) &=&\sum_{\lambda }\frac{3n_{\lambda }}{8}\boldsymbol{%
\hat{d}}_{2}\cdot \boldsymbol{\hat{d}}_{1}\left[ \frac{\sin n_{\lambda
}k_{0}R}{n_{\lambda }k_{0}R}+\frac{\cos n_{\lambda }k_{0}R}{\left(
n_{\lambda }k_{0}R\right) ^{2}}-\frac{\sin n_{\lambda }k_{0}R}{\left(
n_{\lambda }k_{0}R\right) ^{3}}\right]  \nonumber \\
&&-\sum_{\lambda }\frac{3n_{\lambda }}{8}\boldsymbol{\hat{d}}_{2}\cdot
\boldsymbol{\hat{R}\hat{R}}\cdot \boldsymbol{\hat{d}}_{1}\left[ \frac{\sin
n_{\lambda }k_{0}R}{n_{\lambda }k_{0}R}+3\frac{\cos n_{\lambda }k_{0}R}{%
\left( n_{\lambda }k_{0}R\right) ^{2}}-3\frac{\sin n_{\lambda }k_{0}R}{%
\left( n_{\lambda }k_{0}R\right) ^{3}}\right]  \nonumber \\
&&+\sum_{\lambda }\frac{3n_{\lambda }}{8}s_{\lambda } \left( \boldsymbol{%
\hat{d}}_{2}\times \boldsymbol{\hat{d}}_{1}\right) \cdot \boldsymbol{\hat{R}}%
\left[ \frac{\cos n_{\lambda }k_{0}R}{n_{\lambda }k_{0}R}-\frac{\sin
n_{\lambda }k_{0}R}{\left( n_{\lambda }k_{0}R\right) ^{2}}\right],
\label{F1}
\end{eqnarray}%
and
\begin{eqnarray}
F_{2}\left( R\right) &=&\sum_{\lambda }\frac{3n_{\lambda }}{8}\boldsymbol{%
\hat{d}}_{2}\cdot \boldsymbol{\hat{d}}_{1} \left[ \frac{\cos n_{\lambda
}k_{0}R}{n_{\lambda }k_{0}R} -\frac{\sin n_{\lambda }k_{0}R}{\left(
n_{\lambda }k_{0}R\right) ^{2}} -\frac{\cos n_{\lambda }k_{0}R}{\left(
n_{\lambda }k_{0}R\right) ^{3}} \right]  \nonumber \\
&&-\sum_{\lambda }\frac{3n_{\lambda }}{8}\boldsymbol{\hat{d}}_{2}\cdot
\boldsymbol{\hat{R}\hat{R}}\cdot \boldsymbol{\hat{d}}_{1} \left[ \frac{\cos
n_{\lambda }k_{0}R}{n_{\lambda }k_{0}R} -3\frac{\sin n_{\lambda }k_{0}R}{
\left( n_{\lambda }k_{0}R\right) ^{2}} -3\frac{\cos n_{\lambda }k_{0}R}{
\left( n_{\lambda }k_{0}R\right) ^{3}} \right]  \nonumber \\
&&-\sum_{\lambda }\frac{3n_{\lambda }}{8}s_{\lambda } \left( \boldsymbol{%
\hat{d}}_{2}\times \boldsymbol{\hat{d}}_{1}\right) \cdot \boldsymbol{\hat{R}}%
\left[ \frac{\sin n_{\lambda }k_{0}R}{n_{\lambda }k_{0}R} +\frac{\cos
n_{\lambda }k_{0}R}{\left( n_{\lambda }k_{0}R\right) ^{2}} \right]  \nonumber
\\
&&-\sum_{\lambda }\frac{3n_{\lambda }}{8}s_{\lambda } \left( \boldsymbol{%
\hat{d}}_{2}\times \boldsymbol{\hat{d}}_{1}\right) \cdot \boldsymbol{\hat{R}}%
\frac{2}{\pi } \left[ \frac{1}{n_{\lambda }k_{0}R}I_{1} \left(n_{\lambda
}k_{0}R\right)+\frac{1}{\left( n_{\lambda }k_{0}R\right) ^{2}}I_{2}
\left(n_{\lambda }k_{0}R\right)\right].  \label{F2}
\end{eqnarray}
We have defined $\Gamma _{0} \equiv \left(k_{0}^{3}
d^{2}\right)/\left(3\hbar \epsilon _{0}\pi\right) , $ which is the
spontaneous radiation rate of a single dipole in free space. $%
k_{0}=\omega_{0}/c$ where $\omega_{0}$ is the resonance transition angular
frequency of the dipoles and $c$ is the vacuum light speed. $d = \left\vert
\boldsymbol{d}_{a} \right\vert $ is the electric transition dipole momentum,
and $\epsilon _{0}$ is the vacuum permittivity. $\boldsymbol{\hat{d}}_{a} =
\boldsymbol{d}_{a} / d$ are unit vectors that represent the polarization
orientation of dipoles. The interdipole distance $R = \left\vert \boldsymbol{%
r}_{1}-\boldsymbol{r}_{2}\right\vert$, and the unit vector $\boldsymbol{\hat{%
R}} = \left(\boldsymbol{r}_{1}-\boldsymbol{r}_{2}\right)/R $ represents the
interdipole axis. The two functions $I_{1}$ and $I_{2}$ in Eq. (\ref{F2})
are improper integrals defined by
\begin{eqnarray}
I_{1} \left(n_{\lambda }k_{0}R\right)=\int_{0}^{\infty }\frac{\tilde{\xi}%
^{3}e^{-\xi n_{\lambda }k_{0}R}}{\tilde{\xi}^{2}+1}d\tilde{\xi},~~ I_{2}
\left(n_{\lambda }k_{0}R\right)=\int_{0}^{\infty }\frac{\tilde{\xi}^{2}e^{-%
\tilde{\xi}n_{\lambda }k_{0}R}}{\tilde{\xi}^{2}+1}d\tilde{\xi}.
\label{integral2}
\end{eqnarray}
Up to here, the analytical expressions of $A_{L}$ and $A_{T}$ have been
calculated. Associated with Eqs. (\ref{cs}) - (\ref{Ei}), the behaviors of
the dipole pair can be fully described. According to Eq. (\ref{cs}), we
obtain the damping rates of DES and DEAS states
\begin{equation}
\gamma_{\pm} / \Gamma_{0} = \sum_{\lambda} \frac{ n_{\lambda} }{4} \pm
F_{1}\left( R \right),  \label{dr}
\end{equation}
as well as the level shifts due to the interdipole interaction
\begin{equation}
\delta_{\pm} / \Gamma_{0} = \sum_{\lambda }\frac{n_{\lambda }}{2\pi }
\int_{0}^{\infty }d\tilde{\xi}\frac{\tilde{\xi}^{4}}{\tilde{\xi}^{2}-1} \pm
F_{2}\left( R \right).  \label{ls}
\end{equation}

\section{Discussion}

\label{discussion}

\subsection{The behaviors of a single dipole}

Considering a situation that $R \rightarrow \infty$, then $A_{T} \rightarrow
0$ and $C_{1}\left(t\right)=\exp\left(A_{L} t\right)$ while $%
C_{2}\left(t\right)=0$. The physical picture of this result is that dipole 1
spontaneously emits photon while dipole 2 can never be excited. When the two
dipole are very far away separated, they do not interact with each other and
the population ($\left \vert C_{1} \right \vert ^{2}$) on the excited state
of dipole $1$ (the initially excited dipole) decays in a rate of $%
\sum_{\lambda} \Gamma_{0} n_{\lambda}/ 2$. That is why we say the single
dipole mechanism is decided by $A_{L}$. By defining the average refractive
index $\bar{n} = \sum_{\lambda} n_{\lambda}/2, $ we find the spontaneous
radiation rate of one single dipole immersed in optically active media is $%
\gamma_{0} = \bar{n} \Gamma_{0}. $ Similarly, the imaginary part of $A_{L}$
is the vacuum Lamb shift of a single dipole, which is obviously divergent
according to Eq. (\ref{al}). This divergent integral can be renormalized by
Bethe's method
\begin{equation}
\int_{0}^{\infty }d\tilde{\xi}\frac{\tilde{\xi}^{4}}{\tilde{\xi}^{2}-1}
\rightarrow \frac{1}{2} \int_{0}^{m_{e}c/\hbar k_{0}} d\tilde{\xi}\frac{1}{%
\tilde{\xi}-1} =\frac{1}{2} \ln \frac{m_{e}c}{\hbar k_{0}},  \label{renorm}
\end{equation}
in which $m_{e}$ is the observable mass of an electron. Using the expression
of Eq. (\ref{renorm}), we get the vacuum Lamb shift of a single dipole in
optically active medium $\delta_{Lamb}=\bar{n} \Gamma_{0} \ln
\left(m_{e}c/\hbar k_{0}\right)/\left(2 \pi\right). $ $\delta_{Lamb}$ is
position independent, which means it will not lead to any mechanical force.
For this reason, we will not discuss the single dipole vacuum Lamb shift in
the follows.

$A_{L}$ represents the interaction between one single dipole and the vacuum
electromagnetic field. According the analysis above, the radiation rate and
level shift of a single dipole depend on the average rather then the
difference of the refractive indexes for the two circularly polarized
components, and the latter one decides the specific rotation. For this
reason, we can say that the behaviors of a single dipole are affected by the
average refractive indexes rather than the optical rotation of the
surroundings.

\subsection{The collective radiation rates and level shifts}

According to Eqs. (\ref{dr}) and (\ref{ls}), we have obtained the damping
rates $\left( \gamma_{\pm} / \Gamma_{0}= \frac{1}{2} \bar{n} \pm F_{1}
\right)$ as well as the level shifts $\left( \delta_{\pm} / \Gamma_{0}=
\delta_{Lamb}/\Gamma_{0} \pm F_{2} \right)$ of DES and DEAS states. The
total damping rates (and level shifts) are the single dipole part $A_{L}$
plus or minus the collective part $A_{T}$. The value of $A_{T}$ is decided
by the interaction between the dipoles via exchanging virtual photons, therefore it
describes the collective behaviors of the dipole pair. By checking Eqs.~(\ref{F1}) and (\ref{F2}), the most notable properties is that in optically active
surroundings. The cross product ($\boldsymbol{\hat{d%
}}_{1}\times \boldsymbol{\hat{d}}_{2}$) of dipole 1 and dipole 2 contributes to the interdipole
interaction, while this term does not appear in optically inactive case.
This result is not difficult to understand. Imagining one situation that the
orientations of the two dipoles are perpendicular to each other. If the
dipoles are located in optically inactive surroundings, when one virtual
photon are emitted by dipole 1, the polarization of the photon is also
exactly perpendicular to dipole 2 and it is impossible to be absorbed.

For the two dipoles can just interact with each other by exchanging virtual
photons, they can not ``feel" each other in this perpendicular orientation
situation. But when the dipoles are immersed in optically active
surroundings, the polarization of the virtual photon emitted by the donor
dipole will be rotated in propagation, and it may has a component parallel
to the acceptor dipole then the photon can be absorbed. This effect make
term ($\boldsymbol{\hat{d}}_{1}\times \boldsymbol{\hat{d}}_{2}$) appears in
Eqs.~(\ref{F1}) and (\ref{F2}), which respectively represent two-dipole
collective radiation rate and level shift. Furthermore, according to Eqs.~(%
\ref{F1}) and (\ref{F2}), the two circularly polarized components contribute
differently to the interdipole interaction because of the different
refractive indexes.

\subsection{Numerical examples}

To visually show the results given above, we will give out three numerical
examples in the follows. We plot the damping rates and the level shifts as
functions of interdipole distance and compare the different behaviors of the
dipole pair in optically active and inactive surroundings. To specify the
parameters, we set the average refractive index of the surrounding medium to
be $3$ , and the specific rotation divided by wave vector $k$ to be $-1.5$.

\begin{figure}[h]
\centering
\includegraphics[width=0.70\textwidth]{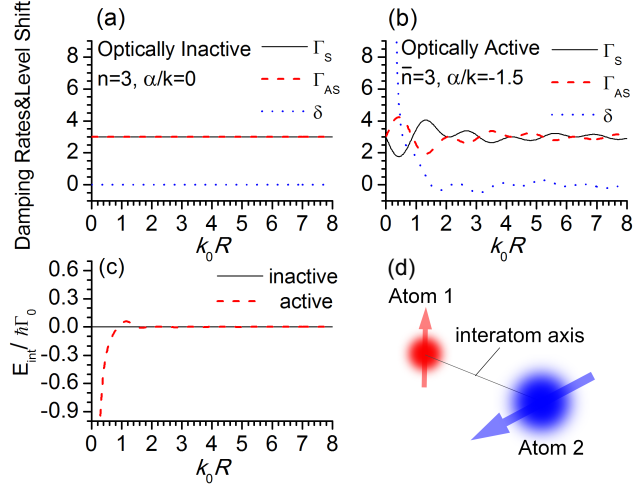}
\caption{(Color online) The polarizations of the two atom are orthogonal and
respectively perpendicular to the interdipole axis, as shown in (d). $%
\Gamma_{S} = 2 \protect\gamma_{+} / \Gamma_{0}$ represents the damping rate
of the population of DES state defined in Eq. (\protect\ref{symstate}) while
$\Gamma_{AS} = 2 \protect\gamma_{-} / \Gamma_{0}$ refers to DEAS state. In
the optically (a) inactive case, the level shift is a constant which means
the two dipoles do not interaction with each other. When the dipoles are
immersed in (b) optically active medium, the RI can be built up by exchanges
of virtual photons between the two dipoles. Furthermore, the dimensionless
dipole-dipole RI $E_{int}/\hbar \Gamma_{0}$ at time $\Gamma_{0}t=1$ are
plotted as functions of interdipole distance in (c). $\protect\delta = \left(%
\protect\delta_{+} - \protect\delta_{-} \right) / \Gamma_{0}$ represents the
dimensionless energy level difference between the DES and DEAS states, while
these two states are degenerate without interdipole interaction. }
\label{chuizhi}
\end{figure}

We first consider a situation that the two dipoles are orthogonally
polarized, and the orientations of them are perpendicular to the interdipole
axis, as shown in Fig. \ref{chuizhi}. According to the physical analysis above, two orthogonally polarized dipoles in vacuum can not interact with each other while an optically active surroundings can help in establishing interdipole RI.
To check this analysis, we plot the level shifts as well as the radiation rates in
optically inactive and active surroundings. The results in the different
situations are compared in Fig. \ref{chuizhi}.
In the optically inactive
case, according to Fig. \ref{chuizhi}(b), the level shift $\delta$ is zero,
which means the DES and DEAS states are degenerate and the two dipoles do
not interact with each other. The damping rates of DES and DEAS state are $3
\Gamma_{0}$, which indicate that dipole $1$ can not feel the existence of
dipole $2$ and just decays alone. In the optically active case, the damping
rates and level shifts vary with the interdipole distance in Fig. \ref%
{chuizhi}(c), which implies an RI between the dipoles. We plot the
interdipole RI in the case of perpendicular dipole orientations as functions
of interdipole distance in Fig. \ref{chuizhi}(c). For RI decays with time,
we choose a time at $\Gamma_{0}t=1$, and use $\hbar \Gamma_{0}$ as the unit
of interaction energy. The results of this numerical experiment agree with
the physical analysis above, and typically shows that optically active
surroundings can noticeably affect interdipole RI.

\begin{figure}[h]
\centering
\includegraphics[width=0.70\textwidth]{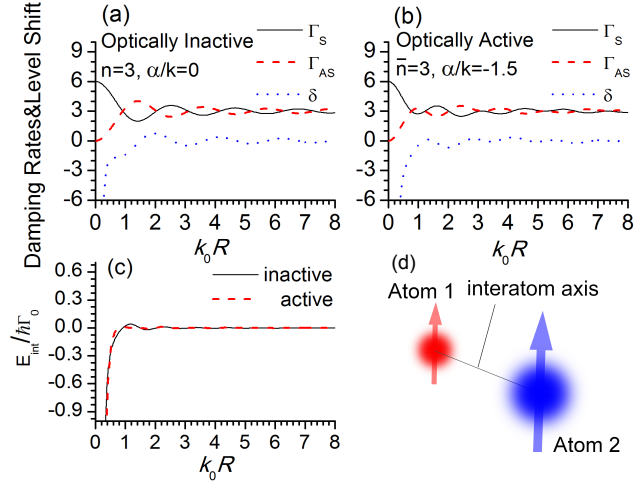}
\caption{(Color online) The polarizations of the two atom are syntropic and
respectively perpendicular to the interdipole axis as shown in (d). The
variations of collective radiation rates and level shift with respect to
interdipole distance $R$ are noticeably different in optically (a) inactive
and (b) active cases. The dimensionless dipole-dipole RI $E_{int}/\hbar
\Gamma_{0}$ at time $\Gamma_{0}t=1$ are plotted as functions of interdipole
distance in (c). RI behaves rather differently in optically active and
inactive surroundings. }
\label{pingxin}
\end{figure}

Now we further consider a situation that the polarizations of the two
dipoles are syntropic and perpendicular to the interdipole axis, as shown in
Fig. \ref{pingxin}.
The level shifts and the radiation rates in optically inactive and active
cases are compared in Fig. \ref{pingxin}(a) and (b). We see that for the
optically inactive case, the first point of minimum of the level shift $%
\delta$ appears in the neighbourhood of $k_{0}R = 3$ while it does near $%
k_{0}R = 2$ for the optically active case. Obviously that the curves in Fig. %
\ref{pingxin}(a) and Fig. \ref{pingxin}(b) are different, which means the
resonance interaction of the dipoles are affected by the optical rotation of
the medium. The dimensionless interaction energy $E_{int}/\hbar\Gamma_{0}$
in the case of syntropic dipole orientations at time $\Gamma_{0}t=1$ are
plotted in Fig. \ref{pingxin}(c).

\begin{figure}[h]
\centering
\includegraphics[width=0.70\textwidth]{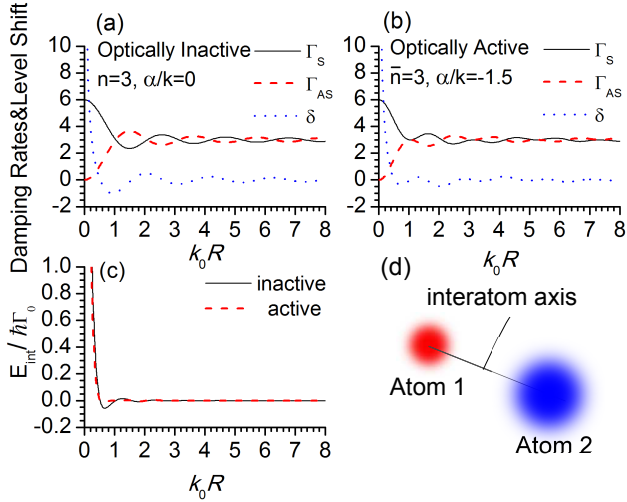}
\caption{(Color online) The polarizations of the dipoles are isotropic ($%
d_{x}=d_{y}=d_{z}=d/\protect\sqrt{3}$) as shown in (d). The variations of
collective radiation rates and level shift with respect to interdipole
distance $R$ are noticeably different in optically (a) inactive and (b) active
cases. The dimensionless dipole-dipole RI $E_{int}/\hbar \Gamma_{0}$ at time
$\Gamma_{0}t=1$ are plotted as functions of interdipole distance in (c). RI
behaves rather differently in optically active and inactive surroundings.}
\label{isotropic}
\end{figure}

Next we consider a situation that the dipoles are isotropically polarized,
namely $d_{x}=d_{y}=d_{z}=d/\sqrt{3}$, as shown in Fig. \ref{isotropic}.
The level shifts and the radiation rates in optically inactive and active
cases are plotted in Fig. \ref{isotropic}(a) and (b), which behave
differently. For example, when the interdipole separation is about $k_{0}R =
1.5 $, the DEAS state decays faster in optically inactive surroundings while
the DES state does in optically active case. The dimensionless interaction
energy $E_{int}/\hbar\Gamma_{0}$ in the case of isotropic polarization at
time $\Gamma_{0}t=1$ are plotted in Fig. \ref{isotropic}(c).

These three numerical examples typically show the effect on RI of optically
active surroundings. In studying RI or using it to develop measurement
methods, the optical rotation of surroundings in which dipoles are immersed
can not be ignored.

\section{Conclusion}

\label{conclusion} In this paper, we have calculated the RI between two
electric dipoles immersed in absorption free optically active surroundings.
The analytical expressions of the collective radiation rates and level
shifts are given out, which satisfy general dipole orientations and interdipole distances. Considering the optical rotation of surrounding
medium, the cross product of two electric dipole moment
appears while this term does not contribute to interdipole RI
in optically inactive case. By comparing the RI in optically inactive and active cases numerically, we obtain that the interdipole RI are noticeably affected by the optical rotation of surroundings. Especially, in the case of that the orientations of
two dipoles are orthogonal, the optical rotation of surrounding medium can
help in building up interdipole RI while there is no interaction in
situation of free space. Our result can help in studying the nature of RI
between electric dipoles immersed in optically active medium, and may
contribute in developing accurate methods of measurement in chemistry and
biology.

\section*{Acknowledgements}

This work was supported by the National Natural Science Foundation of China
under Grant Nos.~11474095, ~11274118, ~11234003, and~11129402, and the
Fundamental Research Funds for the Central Universities.
\appendix

\section{Appendix}

To get Eq. (\ref{c1c2}), we make use of Schr\"{o}dinger equation
\begin{equation}
i\hbar\frac{\partial\left\vert \psi\right\rangle }{\partial t}=\hat {H}%
\left\vert \psi\right\rangle ,  \label{Schordinger}
\end{equation}
to get the equations of motion of the probability amplitudes $C_{i}$ and $%
D_{i, \boldsymbol{k}\lambda }$
\begin{eqnarray}
\dot{C}_{1}=\frac{i}{\hbar }\sum_{a=1,2}\sum_{i^{\prime }=3,4}\sum_{
\boldsymbol{k}\lambda } D_{i^{\prime },\boldsymbol{k}\lambda }A_{k}%
\boldsymbol{d}_{1i^{\prime}}^{a} \cdot \boldsymbol{W}\left( \boldsymbol{k}
\lambda ,\boldsymbol{r}_{a}\right),  \label{c1} \\
\dot{C}_{2}=\frac{i}{\hbar }\sum_{a=1,2}\sum_{i^{\prime }=3,4}\sum_{
\boldsymbol{k}\lambda } D_{i^{\prime },\boldsymbol{k}\lambda }A_{k}%
\boldsymbol{d}_{2i^{\prime}}^{a} \cdot \boldsymbol{W}\left( \boldsymbol{k}%
\lambda ,\boldsymbol{r}_{a}\right),  \label{c2} \\
\dot{D}_{3,\boldsymbol{k}\lambda }= -i\left( \omega _{k}+\omega _{0}\right)
D_{3,\boldsymbol{k}\lambda } +\frac{i}{\hbar }\sum_{a=1,2}\sum_{i^{\prime
}=1,2} C_{i^{\prime }}A_{k}^{\ast }\boldsymbol{d}_{3i^{\prime}}^{a} \cdot
\boldsymbol{W}^{\ast }\left( \boldsymbol{k}\lambda ,\boldsymbol{r}%
_{a}\right),  \label{d3} \\
\dot{D}_{4,\boldsymbol{k}\lambda }= -i\left( \omega _{k}-\omega _{0}\right)
D_{4,\boldsymbol{k}\lambda } +\frac{i}{\hbar }\sum_{a=1,2}\sum_{i^{\prime
}=1,2} C_{i^{\prime }}A_{k}^{\ast }\boldsymbol{d}_{4i^{\prime}}^{a} \cdot
\boldsymbol{W}^{\ast }\left( \boldsymbol{k}\lambda ,\boldsymbol{r}%
_{a}\right).  \label{d4}
\end{eqnarray}%
To solve these equations, we shall first calculate the formal solutions of
Eqs. (\ref{d3}) and (\ref{d4})
\begin{eqnarray}
D_{3,\boldsymbol{k}\lambda }\left( t\right) =\frac{i}{\hbar }A_{k}^{\ast
}\sum_{a=1,2}\sum_{i^{\prime }=1,2}\boldsymbol{d}_{3i^{\prime }}^{a}\cdot
\boldsymbol{W}^{\ast }\left( \boldsymbol{k}\lambda ,\boldsymbol{r}%
_{a}\right) C_{i^{\prime }}\left( t\right) \left[ \pi \delta \left( \omega
_{k}+\omega _{0}\right) -\frac{i}{\left( \omega _{k}+\omega _{0}\right) }%
\right],  \label{d3mar} \\
D_{4,\boldsymbol{k}\lambda }\left( t\right) =\frac{i}{\hbar }A_{k}^{\ast
}\sum_{a=1,2}\sum_{i^{\prime }=1,2}\boldsymbol{d}_{4i^{\prime }}^{a}\cdot
\boldsymbol{W}^{\ast }\left( \boldsymbol{k}\lambda ,\boldsymbol{r}%
_{a}\right) C_{i^{\prime }}\left( t\right) \left[ \pi \delta \left( \omega
_{k}-\omega _{0}\right) -\frac{i}{\left( \omega _{k}-\omega _{0}\right) }%
\right],  \label{d4mar}
\end{eqnarray}%
in which the Markov approximation has been used
\begin{equation}
\int_{0}^{t}C_{i^{\prime }}\left( t^{\prime }\right) e^{-i\left( \omega
_{k}-\omega _{0}\right) \left( t-t^{\prime }\right) }dt^{\prime } \approx
C_{i^{\prime }}\left( t\right) \left[ \pi \delta \left( \omega _{k}-\omega
_{0}\right) -\frac{i}{\left( \omega _{k}-\omega _{0}\right) }\right].
\end{equation}%
Substituting Eqs. (\ref{d3mar}) and (\ref{d4mar}) into Eqs. (\ref{c1}) and (%
\ref{c2}), we can get the equations of $C_{1}$ and $C_{2}$
\begin{eqnarray}
\dot{C}_{1}= A_{11}C_{1}\left( t\right) +A_{12}C_{2}\left( t\right), \dot{C}%
_{2}= A_{22}C_{2}\left( t\right) +A_{21}C_{1}\left( t\right),  \label{c2o}
\end{eqnarray}%
and the coefficients $A_{ab}$ is defined by
\begin{eqnarray}
A_{aa} =-\frac{1}{\hbar ^{2}} \sum_{\boldsymbol{k}\lambda }\left\vert
A_{k}\right\vert ^{2} \left\{ G_{aa}\left( \boldsymbol{k}\lambda \right) \pi
\delta \left( \omega_{k}-\omega_{0}\right) -G_{aa}\left( \boldsymbol{k}
\lambda \right) \frac{i}{\left( \omega_{k}-\omega_{0}\right) }-G_{bb}\left(
\boldsymbol{k}\lambda \right) \frac{i}{\left( \omega_{k}+\omega_{0}\right) }
\right\},  \nonumber \\
A_{ab} =-\frac{1}{\hbar ^{2}} \sum_{\boldsymbol{k}\lambda }\left\vert
A_{k}\right\vert ^{2} \left\{ G_{ba}\left( \boldsymbol{k}\lambda \right) \pi
\delta \left( \omega_{k}-\omega_{0}\right) -G_{ba}\left( \boldsymbol{k}
\lambda \right) \frac{i}{\left( \omega_{k}-\omega_{0}\right) }-G_{ab}\left(
\boldsymbol{k}\lambda \right) \frac{i}{\left( \omega_{k}+\omega_{0}\right) }
\right\},  \label{Aij}
\end{eqnarray}%
in which $a \neq b$. The $G$ functions are defined as $G_{aa\left(b\right)}%
\left( \boldsymbol{k}\lambda \right) =\boldsymbol{d}_{a}\cdot \boldsymbol{W}%
^{\ast }\left( \boldsymbol{k}\lambda ,\boldsymbol{r}_{a}\right) \boldsymbol{W%
}\left( \boldsymbol{k}\lambda ,\boldsymbol{r}_{a\left(b\right)}\right) \cdot
\boldsymbol{d}_{a\left(b\right)} $. The dipole transition matrix elements $%
\boldsymbol{d}_{ij}^{a}$ have been denoted as $\boldsymbol{d}_{a}$ for
simplicity. The initial state is $C_{1}=1$ and $C_{2}=0$, we further define
\begin{equation}
A_{L}=A_{11}=A_{22},A_{T}=A_{12}=A_{21},
\end{equation}
and it is not difficult to get solution (\ref{c1c2}) by solving Eq. (\ref%
{c2o}).


\begin{thebibliography}{99}
\bibitem{ri2} D. P. Craig, and T. Thirunamachandran, \textit{Molecular
Quantum Electrodynamics} (Academic Press, 1984).

\bibitem{ri3} H. Margenau, and N. R. Kestner, \textit{Theory of
Intermolecular Forces} (Elsevier, 1969).

\bibitem{ncomms4610} F. T. Rabouw, S. A. den Hartog, T. Senden and A.
Meijerink, ``Photonic effects on the F\"{o}rster resonance energy transfer
efficiency," Nature Communications \textbf{5}, 3610 (2014).

\bibitem{entanmole} G. K. Brennen, I. H. Deutsch, and P. S. Jessen,
``Entanglement and quantum phase transition in the one-dimensional
anisotropic $XY$ model," Phys. Rev. A \textbf{61}, 062309 (2000). 

\bibitem{entanquad} H. Matsueda, \textit{Coherence and Statistics of Photons
and Atoms}, edited by J. Perina (Wiley, New York, 2001). 

\bibitem{lasercool} W. D. Phillips, ``Nobel Lecture: Laser cooling and
trapping of neutral atoms," Rev. Mod. Phys. \textbf{70}, 721 (1998). 

\bibitem{moleform1} C.M. Dion, C. Drag, O. Dulieu, B. Laburthe Tolra, F.
Masnou-Seeuws, P. Pillet, ``Resonant Coupling in the Formation of Ultracold
Ground State Molecules via Photoassociation," Phys. Rev. Lett. \textbf{86}%
(11), 2253 (2001). 

\bibitem{moleform2} J. Vala, O. Dulieu, F. Masnou-Seeuws, P. Pillet, R.
Kosloff, ``Coherent control of cold-molecule formation through
photoassociation using a chirped-pulsed-laser field," Phys. Rev. A \textbf{63%
}(1), 013412 (2000). 

\bibitem{qs1} A. Kalachev, and S. Kr\"{o}ll, ``Coherent control of
collective spontaneous emission in an extended atomic ensemble and quantum
storage," Phys. Rev. A \textbf{74}(2), 023814 (2006). 

\bibitem{qs2} R. H. Dicke, ``Coherence in Spontaneous Radiation Processes,"
Phys. Rev. \textbf{93}, 99 (1954). 

\bibitem{qs3} M. O. Scully, ``Collective Lamb Shift in Single Photon Dicke
Superradiance," Phys. Rev. Lett. \textbf{102}(14), 143601 (2009). 

\bibitem{moledis} J. Zheng, ``Spectroscopy-Based Quantitative Fluorescence
Resonance Energy Transfer Analysis," in \textit{Ion Channels: Methods and
Protocols. Methods in Molecular Biology}, N. Gamper, ed. (Humana Press,
2006). 

\bibitem{tufe} K. Truong, and M. Ikura, ``The use of FRET imaging microscopy
to detect protein--protein interactions and protein conformational changes
in vivo," Current Opinion in Structural Biology \textbf{11}(5), 573 (1991).

\bibitem{dd} H. Volkhard, ``Fluorescence Resonance Energy Transfer," in
\textit{Principles of Computational Cell Biology}, H. Kiaris, ed.
(Wiley-VCH, 2008). 

\bibitem{forster} Th. F\"{o}rster, \textit{Modern Quantum Chemistry}, O.
Sinanoglu, ed. (Academic, 1965), Pt. 3. 

\bibitem{ugfeb} B. A Pollok, and R. Heim, ``Using GFP in FRET-based
applications," Trends in Cell Biology \textbf{9}(2), 57 (1999). 

\bibitem{fretmi} R. B. Sekar, and A. Periasamy, ``Fluorescence resonance
energy transfer (FRET) microscopy imaging of live cell protein
localizations," The Journal of Cell Biology \textbf{160}(5), 629-633 (2003).

\bibitem{dssc} B. O'Regan,and M. Gr\"{a}tzel, ``A low-cost, high-efficiency
solar cell based on dye-sensitized colloidal TiO2 films," Nature \textbf{353}%
, 737-740 (1991). 

\bibitem{ysm} Shu-Ming Yang, \textit{Dye-Sensitized Nanocrystalline
Photovoltaic Solar Cells} (Zhengzhou Univerdity Press, 2007). 
\end{thebibliography}
\end{document}